\documentclass[epjCONF, onecolumn]{svjour}
\usepackage{graphics}
\usepackage[varg]{txfonts} 
\usepackage[latin1]{inputenc}
\def\degr{\hbox{$^\circ$}}
\session-title{Conference Title, to be filled}
\begin{document}
\title{Elemental abundances in AGB stars and the formation of the Galactic bulge}
\author{S.\ Uttenthaler\inst{1}\fnmsep\thanks{\email{stefan.uttenthaler@univie.ac.at}} \and J.\ A.\ D.\ L.\ Blommaert\inst{2} \and T.\ Lebzelter\inst{1} \and N.\ Ryde\inst{3} \and P.\ R.\ Wood\inst{4} \and M.\ Schultheis\inst{5} \and B.\ Aringer\inst{6}}
\institute{University of Vienna, Department of Astronomy, T\"urkenschanzstra\ss e 17, 1180 Vienna, Austria \and Instituut voor Sterrenkunde, K.\ U.\ Leuven, Celestijnenlaan 200D, 3000 Leuven, Belgium \and Department of Astronomy and Theoretical Physics, Box 43, SE-22100 Lund, Sweden \and Research School of Astronomy and Astrophysics, Australian National University, Cotter Rd, Weston Creek ACT 2611, Australia \and Observatoire de Besan\c{c}on, 41bis, avenue de l'Observatoire, 25000 Besan\c{c}on, France \and INAF - Padova Astronomical Observatory, Vicolo dell'Osservatorio 5, 35122, Padova, Italy}

\abstract{We obtained high-resolution near-IR spectra of 45 AGB stars located
  in the Galactic bulge. The aim of the project is to determine key elemental
  abundances in these stars to help constrain the formation history of the
  bulge. A further aim is to link the photospheric abundances to the dust
  species found in the winds of the stars. Here we present a progress report of
  the analysis of the spectra.
}

\maketitle

\section{Introduction}\label{intro}

The bulge of the Milky Way Galaxy is one of its major components, but its
formation is still poorly understood. It still holds many surprises for us,
see e.g.\ the recently discovered double red clump \cite{Nat10,MZ10}. Elemental
abundances in bulge stars are an important tool to constrain formation
scenarios. K/M giants and PNe in the bulge have already been studied for
their abundance patterns, but not the evolutionary state between them, the AGB.
We obtained high-resolution near-IR spectra of 45 AGB stars with the CRIRES@VLT
spectrograph to better characterise the giant star population in the bulge.

\section{The sample}\label{sample}

The CRIRES targets are located at various Galactic latitudes, from $b=-1\degr$
to $b=-12\degr$, but most of them are less than $4\degr$ south of the Galactic
plane. The stars have been selected either from ISOGAL photometry
to cover a range in $K -[15]$ (i.e.\ mass-loss rate), or from the sample of
outer bulge AGB stars that has been observed with UVES by \cite{Utt07}. There
are several indications that suggest an intermediate age of bulge AGB
variables, e.g.\ their pulsation period distribution \cite{GB05} and the
presence of technetium in some of them \cite{Utt07}. For these reasons, it is
interesting to compare their chemical abundances to other populations in the
bulge.

%
%

\section{Analysis}\label{anal}

The target stars were mainly observed in one setting in the H-band and one
setting in the K-band. These regions contain numerous atomic (Fe, Ti, Si,
\dots) and molecular lines (OH, CN, and CO, including $^{13}$CO). They allow for
the measurement of abundances of all the involved elemental species.

\begin{figure}
  \centering
  \resizebox{0.75\columnwidth}{!}{\includegraphics{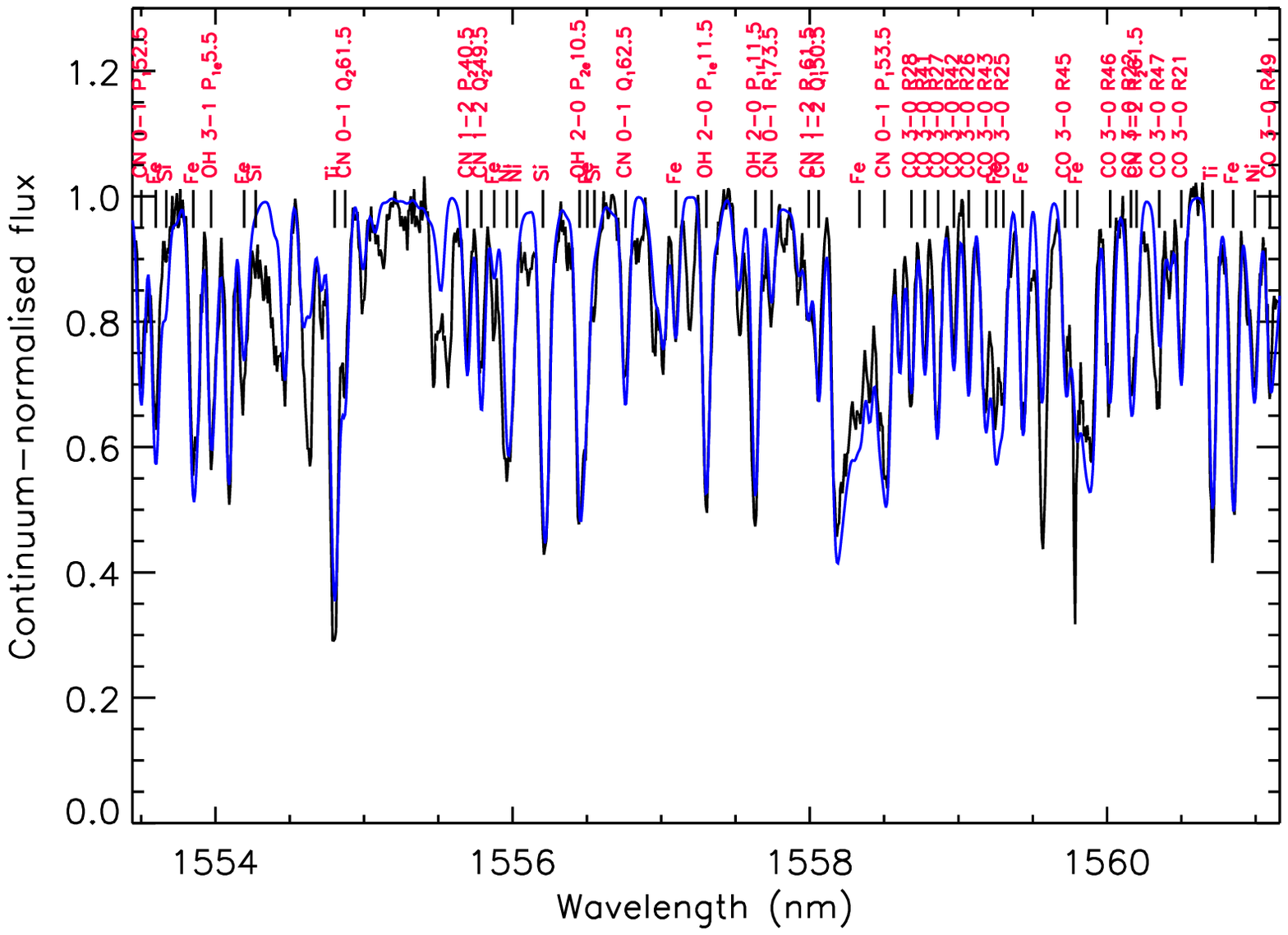}}
  \caption{Spectrum of the star J174924.1-293522 on chip~2 of the H-band
    CRIRES setting (black). The blue graph is a synthetic spectrum based on
    COMARCS models with $T_{\rm eff} = 3700$\,K and $\log g=+0.30$. It was
    interpolated within the 80 grid models to the best-fit overall abundance
    pattern of [Fe/H]=$+0.15$, [$\alpha$/Fe]=--0.36, and C/O=0.48.}
  \label{fig1}
\end{figure}

We employ hydrostatic COMARCS model atmospheres and COMA spectral synthesis
\cite{Ari09} in the analysis of the spectra. Stellar parameters (luminosity,
mass, temperature, $\log g$) have been estimated from photometry,
low-resolution optical spectra, and the pulsation periods of the stars. For
every star, a grid of model atmospheres and spectra with 80 different abundance
patterns is calculated. We interpolate between the grid spectra to determine
the best-fitting overall abundance pattern. A $\chi^{2}$ minimisation method is
employed in this step. Of particular interest is the run of [$\alpha$/Fe] vs.\
[Fe/H] because it will give information about the formation time scale of the
population from which the AGB stars descend. Figure~\ref{fig1} shows an example
of an observed spectrum, together with the best-fit interpolated spectrum.


\section{The connection to dust}\label{dust}

We also have Spitzer/IRS spectra of all targets, from which their dominant dust
species are known. A variety of dust feature is found, including amorphous
silicates and aluminum oxide, but also the 13 and 19.5\,${\rm \mu m}$ features
with uncertain carriers \cite{Vanh}. One central aim of the project is to
investigate the connection between photospheric abundances of key elements in
the dust formation (e.g.\ Mg, Al) and the dust species found in the stellar
wind. With our data we will be able to better understand the bulge, but also
the AGB stars themselves\footnote{This research was funded by the Austrian
  Science Fund (FWF): P22911-N16, P21988-N16, and P23737-N16.}!

\end{document}